\documentclass{jps-cp}
\usepackage{cite}

\title{The Physics of UHECRs: Spectra, Composition and the Transition Galactic-Extragalactic}

\author{Roberto \textsc{Aloisio}$^{1,2}$}

\inst{$^{1}$Gran Sastso Science Institute, L'Aquila, Italy \\
$^{2}$INFN - Laboratori Nazionali del Gran Sasso, Assergi (AQ), Italy}

\email{roberto.aloisio@gssi.it}

\recdate{\today}

\abst{We review the experimental evidences about flux and mass composition of ultra high energy cosmic rays in connection with theoretical scenarios concerning astrophysical sources. In this context, we also address the discussion about the expected transition between cosmic rays produced inside the Galaxy and those coming from the intergalactic space.}

\kword{ultra high energy cosmic rays, high energy astrophysics}

\begin{document}
\maketitle

\section{Introduction}
Ultra High Energy Cosmic Rays (UHECR), namely Cosmic Rays with energies larger than $10^{17}\div 10^{18}$ eV, are the most energetic particles observed in nature, with energies exceeding $10^{20}$ eV. The first observation of particles with such extreme energies dates back to the Volcano Ranch experiment in 1962 \cite{Linsley:1963km}. Nowadays, the most advanced experiments to detect UHECR are the Pierre Auger Observatory in Argentina \cite{ThePierreAuger:2015rma}, far the largest experimental setup devoted to the study of UHECR, and the Telescope Array (TA) experiment \cite{AbuZayyad:2012kk,Tokuno:2012mi}, placed in the United States, with roughly $1/10$ of the Auger statistics. 

The experimental study of UHECR clarified few important characteristics of these particles: (i) are charged with a limit on photon and neutrino fluxes around $10^{19}$ eV at the level of few percent and well below respectively \cite{Abraham:2009qb,Abu-Zayyad:2013dii,Abreu:2013zbq}, (ii) the spectra observed at the Earth show a slight flattening at energies around $5\times 10^{18}$ eV (called the ankle) with (iii) a steep suppression at the highest energies \cite{ThePierreAuger:2013eja,Abu-Zayyad:2013qwa}.
 
The composition of UHECR is still matter of debate. Before the advent of Auger the experimental evidences, including the latest observations by TA, were all pointing toward a light composition with a proton dominated flux until the highest energies \cite{Berezinsky:2002nc}. The measurements carried out by the Auger observatory \cite{Aab:2014kda} have shown that the mass composition of CRs, from prevalently light at $\sim 10^{18}$ eV, becomes increasingly heavier towards higher energies. 

Mass composition is inferred from the mean value of the depth of shower maximum $\langle X_{max} \rangle$ and its dispersion (RMS) $\sigma(X_{max})$. The combined analysis of $\langle X_{max} \rangle$ and $\sigma(X_{max})$, even if not conclusive, allows to obtain important indications on the actual mass composition \cite{Aloisio:2007rc,Kampert:2012mx} (for a review see also \cite{Engel:2011zzb} and references therein).

\begin{figure}[!h]
\centering\includegraphics[scale=0.6]{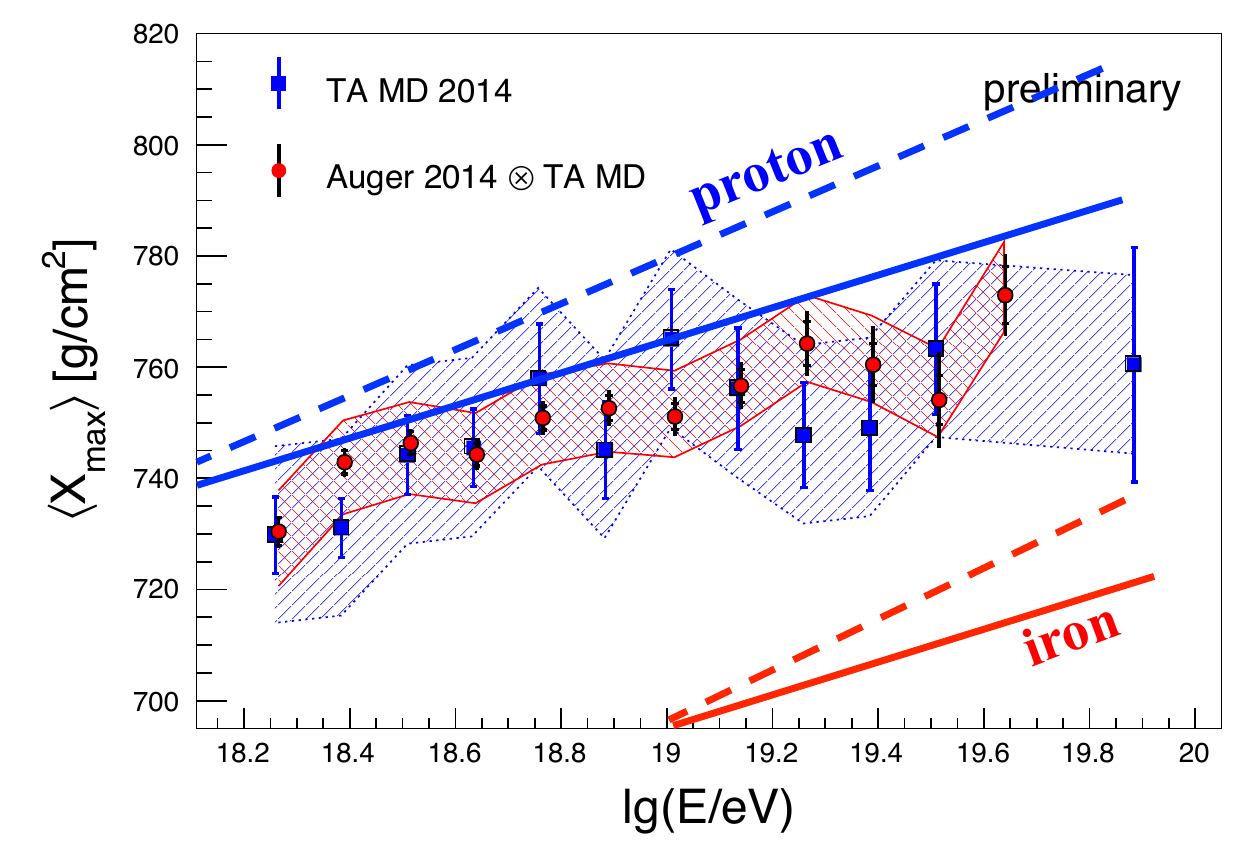}
\includegraphics[scale=0.27]{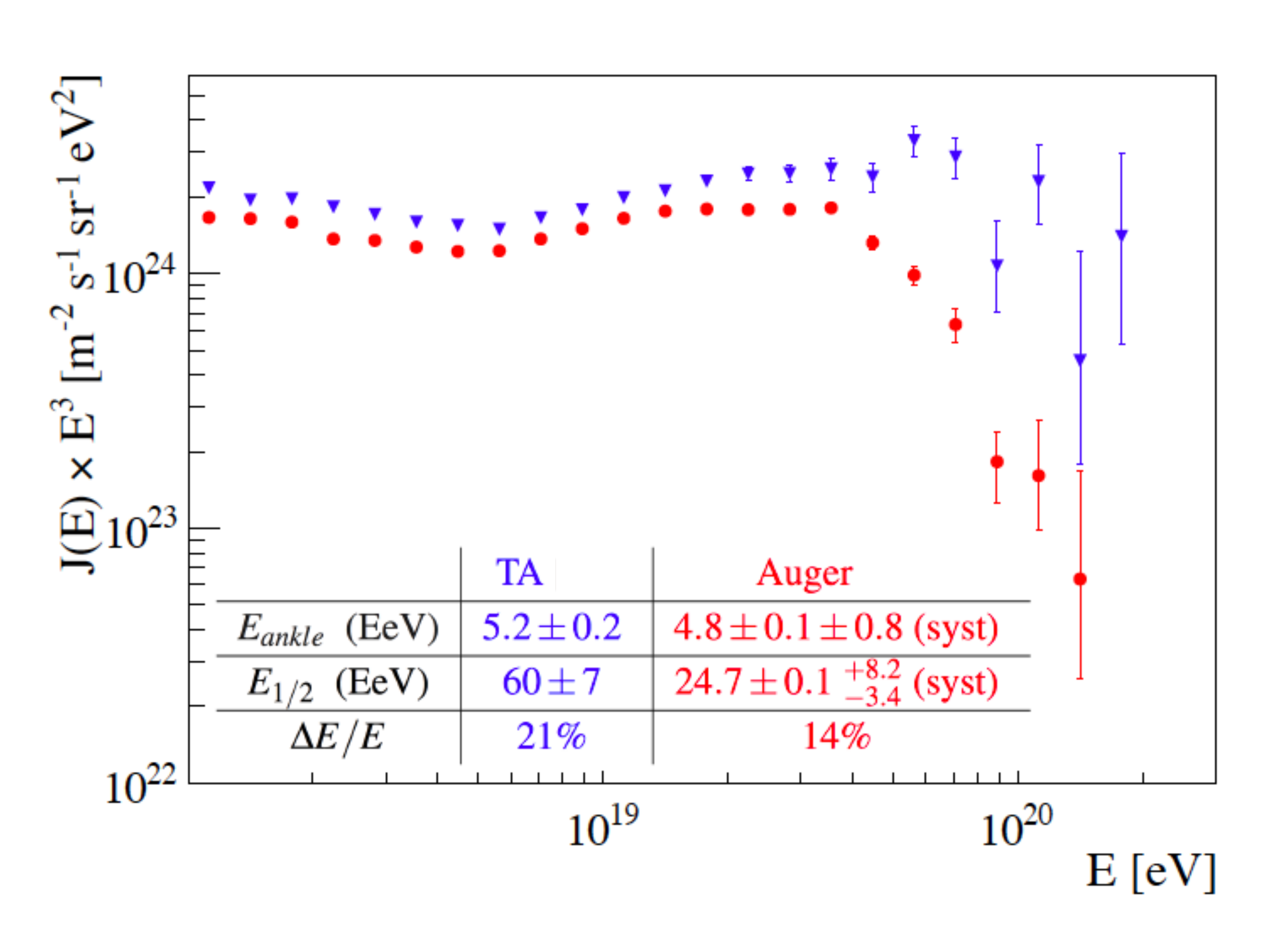}
\caption{[Left panel] Comparison of the $\langle X_{max} \rangle$ as obtained by the Auger-TA working group \cite{Abbasi:2015xga} with superimposed (by the author) the theoretical expectation of pure proton or iron composition (as labeled) in the case of QGS-Jet-II-03 (solid lines) and Sybil 2.1 (dashed lines) hadronic interaction models (curves taken from figure 30 of \cite{Abbasi:2014sfa}). [Right panel] Energy spectra of UHECR as released in 2015 by Auger (red points) and TA (blu points). Also labelled, in both datasets, are: the energy of the ankle $E_{ankle}$, the energy of the high energy suppression $E_{1/2}$ and the systematics in energy determination (figure taken from \cite{Verzi:2015dna}).}
\label{fig:exp}   
\end{figure}

A joint working group made of members of both collaborations, TA and Auger, has recently concluded that the results of the two experiments are not in conflict once systematic and statistical uncertainties have been taken into account \cite{Abbasi:2015xga}. This conclusion is reached by comparing $\langle X_{max} \rangle$ obtained from Auger data, analysed as if taken through the TA's experimental setup, with those obtained through the genuine TA observations. In figure \ref{fig:exp} (left panel) we plot this comparison, as published in \cite{Abbasi:2015xga}, and, superimposed, the theoretical expectations, computed by the author, of a pure proton or iron composition, according to QGS-Jet-II-03 (solid lines) and Sybil 2.1 (dashed lines) hadronic interaction models (curves taken from figure 30 of \cite{Abbasi:2014sfa}). The result of the working group Auger-TA, though encouraging on one hand, is not conclusive and casts serious doubts on the possibility of reliably measuring the mass composition at the highest energies, unless some substantially new piece of information becomes available. 

Comparing the two datasets of Auger and TA, it should be also noted that the spectra measured by the two experiments, though being in general agreement, differ beyond the systematic error at the highest energies (where mass composition differs the most) in such a way that TA claims a spectral suppression at $\gtrsim 5\times 10^{19}$ eV while Auger shows the suppression at sensibly lower energies \cite{Verzi:2015dna}. In figure \ref{fig:exp} (right panel) we plot this comparison as presented at the last International Cosmic Rays Conference \cite{Verzi:2015dna}, also labelled are the position of the ankle and the high energy suppression as measured in the two datasets. Apart from a shift in the energy determination, which can be reabsorbed in systematics uncertainties as labelled in right panel of figure \ref{fig:exp}, it follows that the two datasets seem discrepant at the highest energies in both shape and position of the suppression. 

In the present paper we will address the main theoretical ideas and models aiming at describing the experimental evidences briefly summarised above. The paper is organised as follows: in section \ref{sec:spec} we discuss the experimental evidences of UHECRs in connection with the details of particles transport, in section \ref{sec:trans} we address the transition from galactic to extra-galactic cosmic rays (CR) and, finally, we conclude in section \ref{sec:conclu}. 
  
\section{Transport of UHECRs Spectrum and Mass Composition}
\label{sec:spec}

The extragalactic origin of UHECRs, at least at energies above the ankle $E>10^{19}$ eV, is widely accepted \cite{Aloisio:2012ba}. 
\begin{figure}[!h]
\centering\includegraphics[scale=0.6]{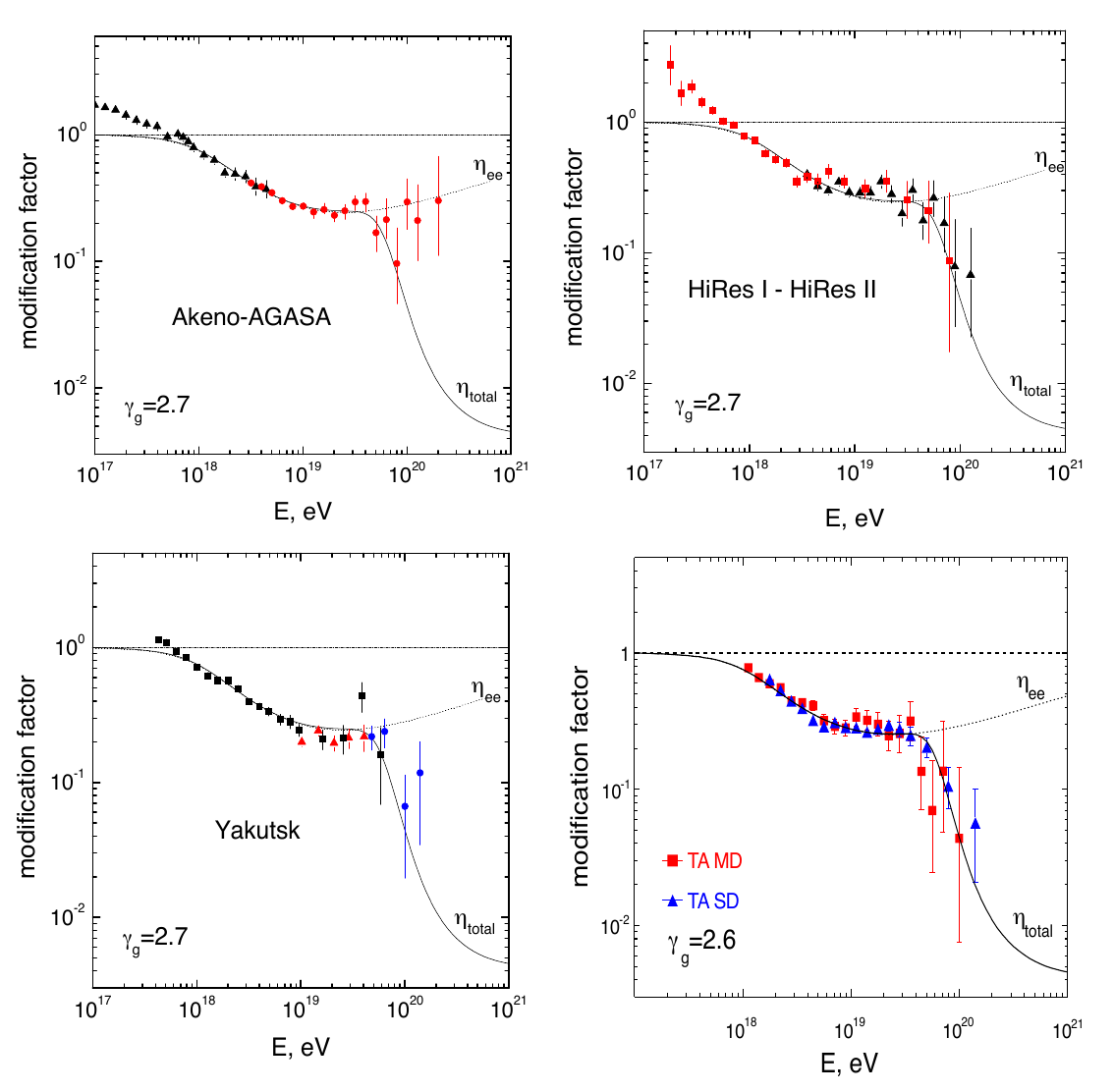}
\includegraphics[scale=0.6]{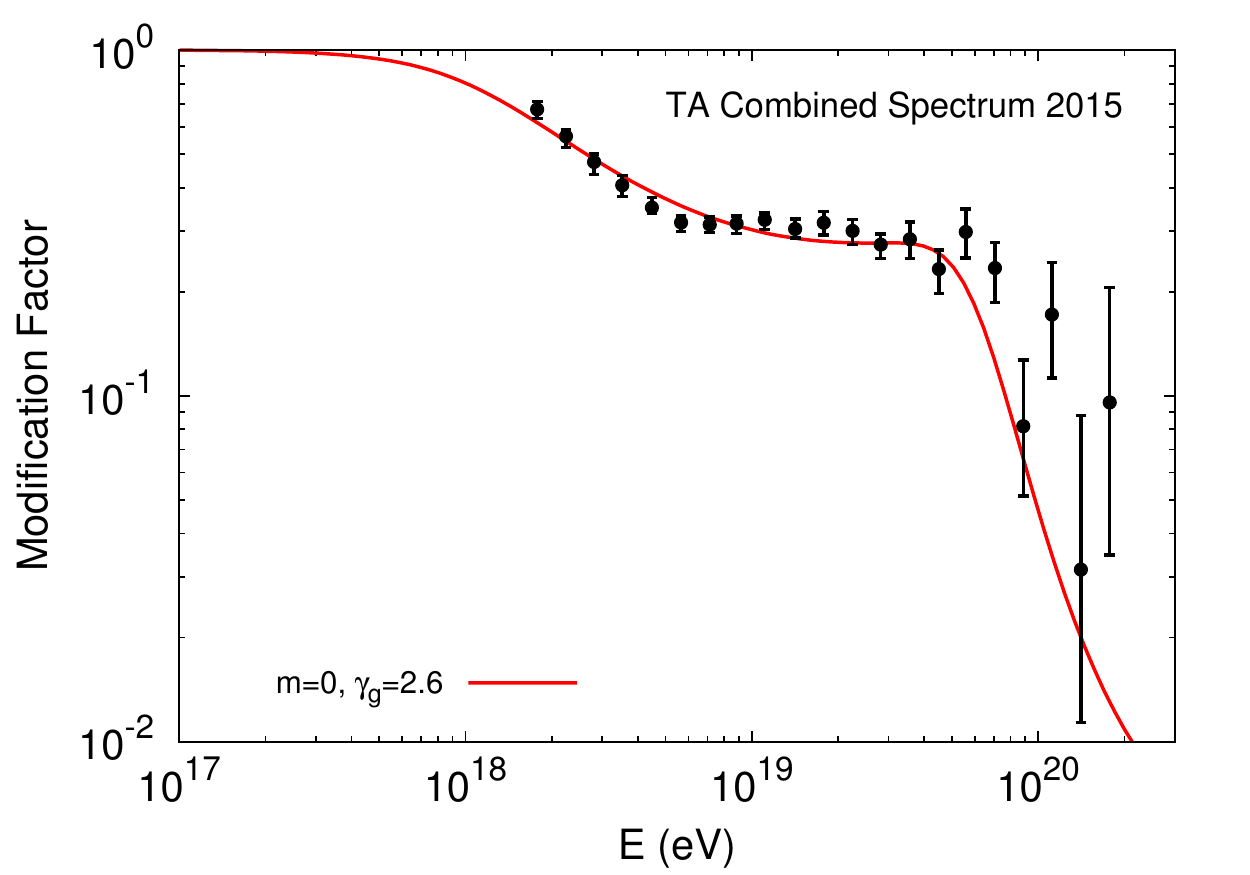}
\caption{[Left panel] Comparison of the modification factor with experimental data before 2006 of different experiments as labeled (see \cite{Berezinsky:2013kfa} and references therein). [Right panel] Comparison of the modification factor with the latest TA data \cite{Ivanov:2015pqx}.}
\label{fig:dip}   
\end{figure}
The propagation of UHECR through the intergalactic medium is conditioned primarily by astrophysical photon backgrounds and, if any, by the presence of magnetic fields. The astrophysical backgrounds involved are the Cosmic Microwave Background (CMB) and the Extragalactic Background Light (EBL). 

The interactions of UHECR (protons or heavier nuclei) with astrophysical backgrounds give rise to the processes of: pair-production, photo-pion production and, only in the case of nuclei heavier than protons, photo-disintegration. Moreover, protons propagation is affected only by the CMB while for nuclei, and only in the case of photo-disintegration, also the EBL plays a role \cite{Aloisio:2008pp,Aloisio:2010he}. 

As discussed above, the observations of Auger and TA are currently not providing an unambiguous measurement of the mass composition hence, in what follows, we will discuss separately the two cases of a pure proton composition (according to TA data) and a mixed composition with heavy nuclei contributing to the flux (according to Auger data). In the following analysis we always assume a power law injection $\propto E^{-\gamma_g}$ of the accelerated particles at the sources. 

In the case of a pure proton composition the only relevant astrophysical background is the CMB \cite{Aloisio:2008pp,Aloisio:2010he}. This fact makes the propagation of UHE protons free from the uncertainties related to the background, being the CMB spectrum a pure black body with an exactly known red-shift evolution. In this case any signature of the propagation in the observed spectrum can be easily referred to the assumptions made at the source, subtracting the effects of the interactions suffered during propagation. In order to isolate these effects it is useful the so-called modification factor $\eta(E)$ defined as the ratio \cite{Berezinsky:2002nc}: $\eta(E)=\frac{J_p(E)}{J_{unm}(E)}$ where $J_p$ is the protons spectrum, computed with all energy losses taken into account, and $J_{unm}(E)$ is the unmodified spectrum computed taking into account only energy losses due to the expansion of the universe. 

\begin{figure}[!h]
\centering\includegraphics[scale=0.5]{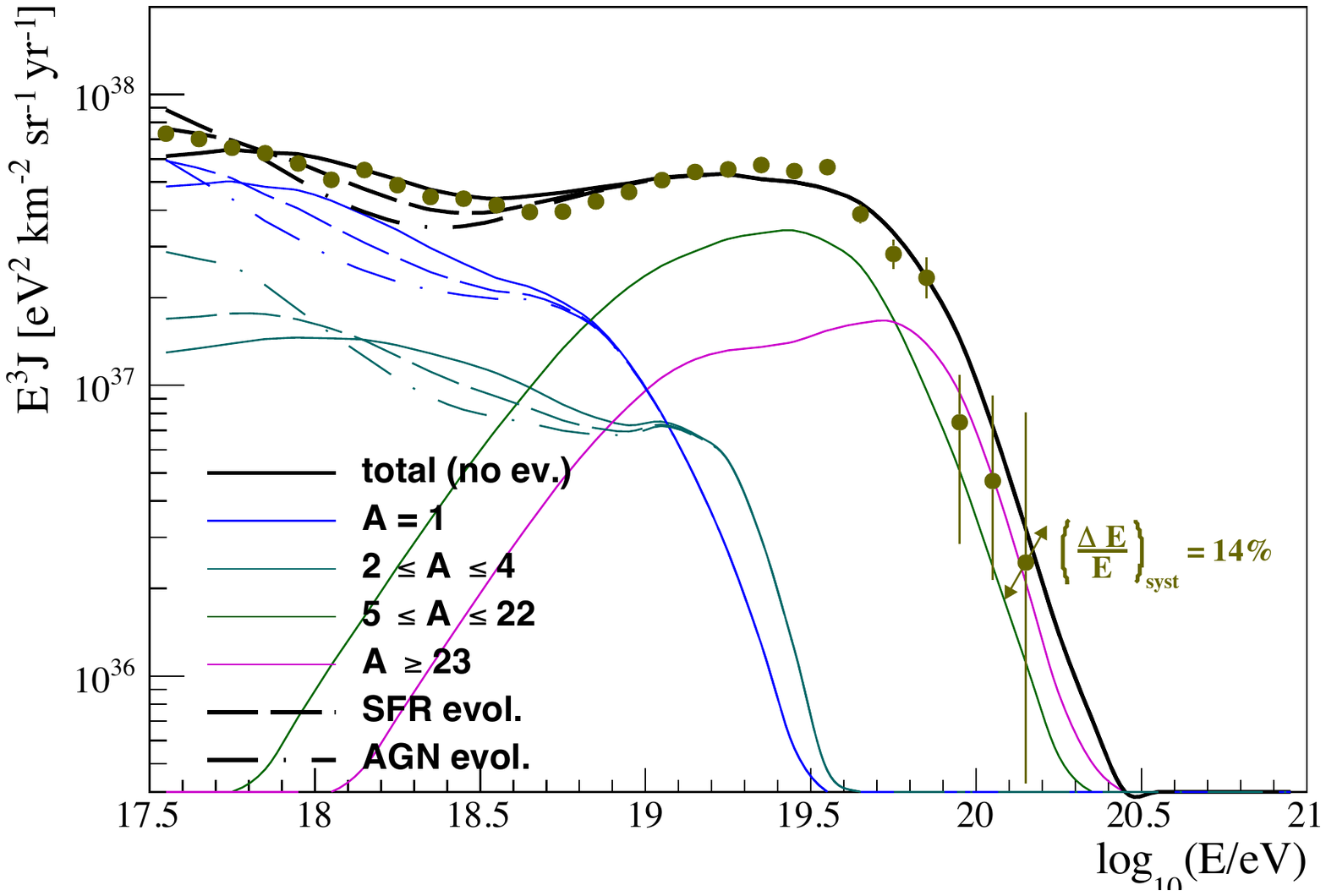}
\begin{minipage}[b]{0.3\textwidth}
\includegraphics[scale=0.25]{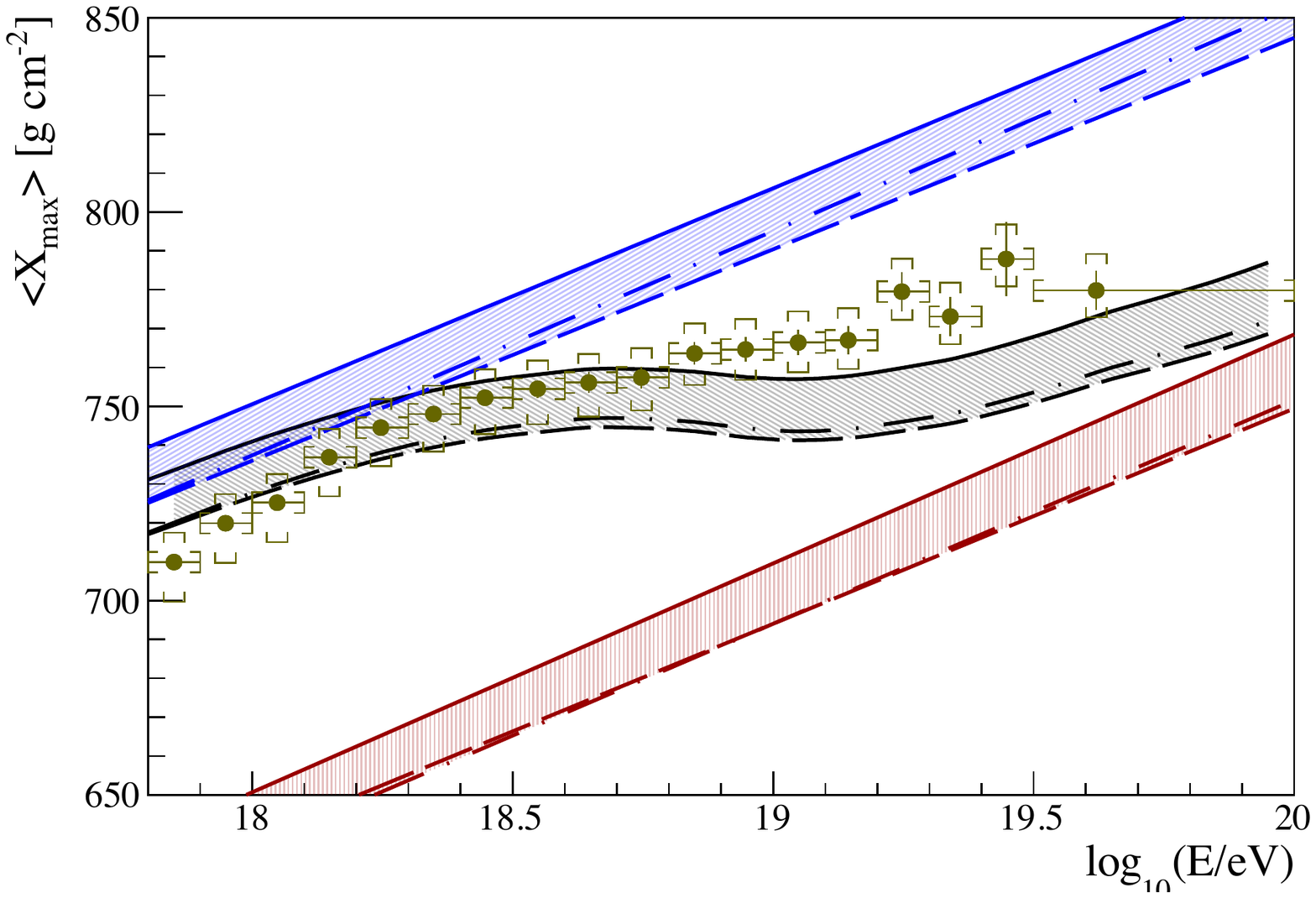}
\includegraphics[scale=0.25]{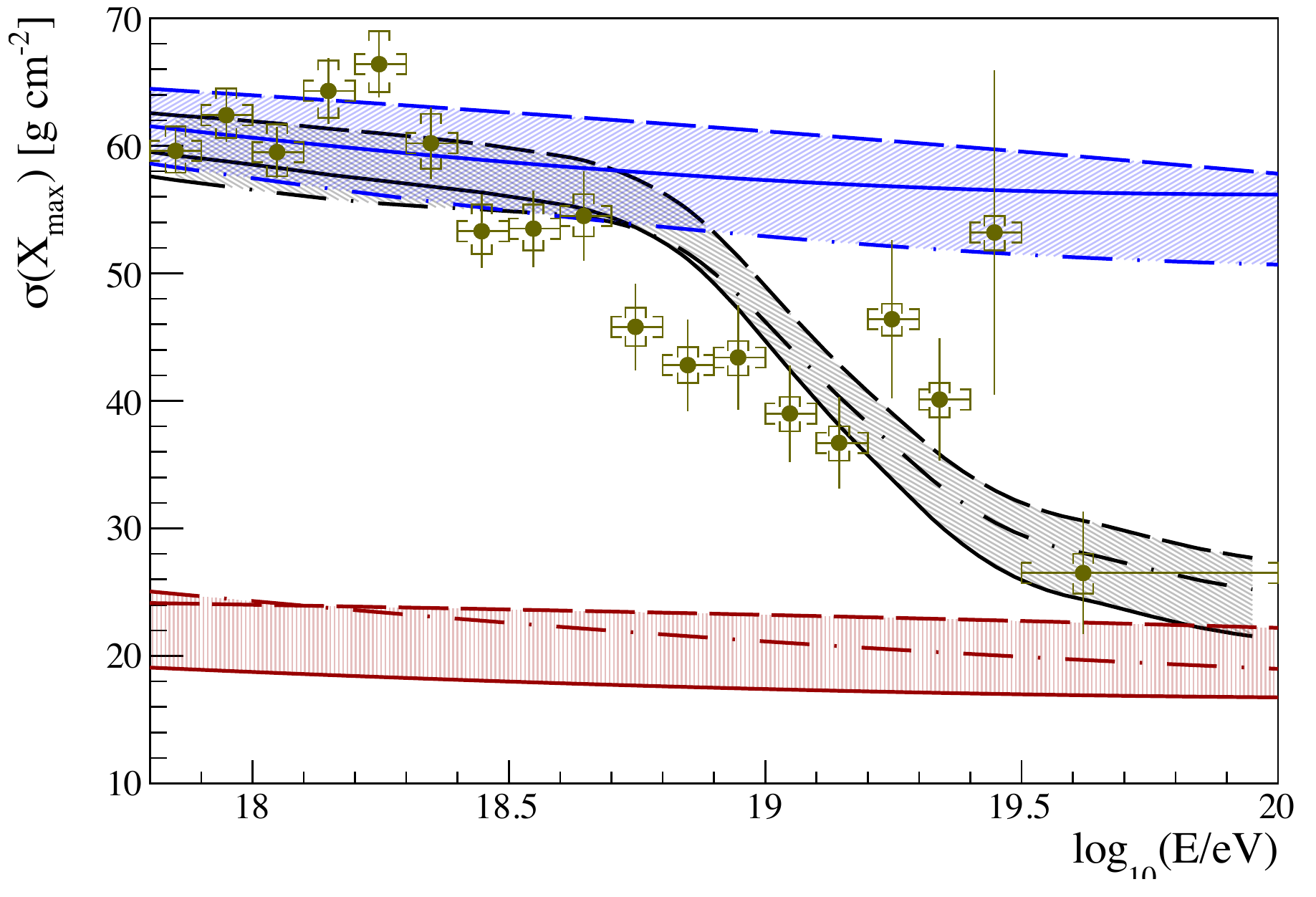}
\end{minipage} 
\caption{Comparison of the Auger observations with theoretical expectations in the case of models with mixed composition. [Left Panel] Model with two classes of sources as in \cite{Aloisio:2015ega}. Continuos, dashed and dot-dashed lines correspond respectively to the cases of: no cosmological evolution of sources, evolution as the star formation rate and as active galactic nuclei. [Right Panel] Comparison of the elongation rate and its root mean square computed assuming the model in the left panel as discussed in \cite{Aloisio:2015ega}.}
\label{fig:JCAP}   
\end{figure}

Particularly relevant is the behaviour associated to the pair-production energy losses that, named "dip" \cite{Berezinsky:2002nc}, reproduces quite well the ankle observed in the spectrum, provided that the injection power law at the source is around $\gamma_g=2.6\div 2.7$. In figure \ref{fig:dip} we plot the theoretical modification factor together with the experimental data of several detectors as labeled, which all claim a pure proton composition. From these figures it is evident that the behaviour of the pair production dip reproduces quite well the ankle observed in the UHECR spectrum. 

The remarkable feature of the dip model is the ability of explaining experimental data with only one extragalactic component of pure protons, directly linking the flux behaviour to the energy losses suffered by propagating particles. From this analysis follows that the required emissivity depends on the injection power law index, which shows a best fit value that ranges from $\gamma_g=2.5$ (for strong cosmological evolution) up to $\gamma_g=2.7$ (without evolution) \cite{Berezinsky:2002nc,Aloisio:2006wv}. 

The discussion above was centred around the hypothesis of a pure proton composition. The qualitative new finding that mass composition might be mixed has served as a stimulus to build models that can potentially explain the phenomenology of Auger data. These models all show that the Auger spectrum and mass composition at $E\ge 5\times 10^{18}$ eV can be fitted at the same time only at the price of requiring very hard injection spectra for all nuclei (with $\gamma_g=1\div 1.6$) and a maximum acceleration energy $E_{max}\le 5 Z\times 10^{18}$ eV. The need for hard spectra can be understood taking into account that the low energy tail of UHECR reproduces the injection power law. Therefore, taking $\gamma\ge 2$ cause the low energy part of the spectrum to be polluted by heavy nuclei thereby producing a disagreement with the light composition observed at low energy. 

One should appreciate here the change of paradigm that these findings imply: while in the case of a pure proton composition it is needed to find sources and acceleration mechanisms able to energise CR protons up to energies larger than $10^{20}$ eV with steep injection ($\gamma_g\simeq 2.5\div 2.7$), the Auger data require that the highest energy part of the spectrum ($E>5\times 10^{18}$ eV) has a flat injection ($\gamma_g\simeq 1.0\div 1.6$) being dominated by heavy nuclei with protons' maximum energy not exceeding few$\times 10^{18}$ eV. By accepting the new paradigm, it follows that the Auger spectrum at energies below $5\times 10^{18}$ eV requires an additional component that, composed by protons and helium nuclei with steep injection $\gamma_g\simeq 2.6\div 2.7$, could be, in principle, of galactic or extra-galactic origin. Nevertheless, the anisotropy expected for a galactic light component extending up to $10^{18}$ eV exceeds by more than one order of magnitude the upper limit measured by Auger \cite{Abreu:2012ybu}. The possible origin of the additional (light and extra-galactic) component can be modelled essentially in two ways: (i) assuming the presence of different classes of sources: one injecting also heavy nuclei with hard spectrum and the other only proton and helium nuclei with soft spectrum \cite{Aloisio:2013hya,Taylor:2013gga} or (ii) identifying a peculiar class of sources that could provide at the same time a steep light component and a flat heavy one \cite{Globus:2015xga,Blasi:2015esa,Unger:2015laa}.

In figure \ref{fig:JCAP} we plot the comparison of Auger data on flux and mass composition with the theoretical expectation in the case of two classes of generic sources as discussed in \cite{Aloisio:2015ega}. In the left panel of figure \ref{fig:JCAP} we also plot the behaviour of the spectra computed with different assumptions on the cosmological evolution of sources \cite{Aloisio:2015ega}: no cosmological evolution (solid line), the evolution typical of the star formation rate (dashed line) and of active galactic nuclei (dot-dashed line). 

In right panel of figure \ref{fig:JCAP}, chemical composition is inferred from the mean value of the depth of shower maximum $\langle X_{max} \rangle$ and its dispersion $\sigma(X_{max})$ (computed as discussed in \cite{Abreu:2013env}). In figure \ref{fig:JCAP} (right panel), to highlight the uncertainties in the atmospheric shower development, four different models of hadronic interaction were included in the coloured bands (see \cite{Aloisio:2015ega} and references therein). 

\section{Transition Galactic-Extragalactic Cosmic Rays} 
\label{sec:trans}

Historically, the transition from galactic CRs to extragalactic CRs has been assumed to take place at the ankle: CR iron nuclei were assumed to be accelerated up to energies in excess of $\sim 10^{19}$ eV , where they would leave room to extragalactic CR protons, injected with spectrum $\sim E^{-2}$. 

This picture remained virtually untouched until it was proposed the dip model discussed above. In this model the galactic CR spectrum is required to end with a heavy composition at much lower energies, $\sim 10^{17}$ eV, two orders of magnitude below the energy of the ankle. This picture is roughly consistent with the idea that galactic super nova remnants may accelerate protons up to the knee. 

A careful investigation of the transition region in the context of the dip and ankle models was carried out (see \cite{Aloisio:2012ba} and references therein), assuming an exponential cutoff in the galactic CR component. The authors concluded that the ankle model is basically ruled out by measurements of the depth of shower maximum, $X_{max}(E)$, and its RMS fluctuations. 

A turning point in the investigation of the transition region is represented by the recent measurement of the spectra of the light and heavy components of CRs in the energy region $10^{17}-10^{18}$ eV by KASCADE-Grande \cite{Apel:2013dga}. These measurements found evidence for an ankle-like feature of the light component at $\sim 10^{17}$ eV and a knee-like feature in the heavy component at roughly the same energy. The former can be interpreted as the signature of the transition to a light extragalactic CR composition, while the latter can be interpreted as the end of the galactic heavy CR component. The main reason to believe that the light KASCADE-Grande component is of extragalactic origin is the low level of anisotropy observed by Auger at $10^{18}$ eV, that seems to be ad odds with a galactic origin of protons at those energies. It is worth stressing that if indeed the light CR component measured by KASCADE-Grande is of extragalactic origin, the transition region becomes weakly dependent upon whether extragalactic CRs are all light (dip model) or have mixed mass composition, since in the energy region $10^{17}-10^{18}$ eV the expected composition is light in both scenarios (see the discussion above). What is less clear is whether this light component reflects a class of sources which are different from those that produce nuclei (with a hard injection spectrum) or rather the light component is due to interaction processes inside the sources, as in \cite{Globus:2015xga,Blasi:2015esa,Unger:2015laa}.

\section{Conclusions}
\label{sec:conclu}

We conclude highlighting the two principal avenues on which the study of UHECR should develop in the near future. From one hand, as discussed above, a firm experimental determination of the mass composition is an unavoidable step forward in this field of research. On the other hand, the highest energy regime, typically the trans-GZK energies $E \gtrsim 5\times 10^{19}$ eV, still remains less probed with not enough statistics to firmly detect possible anisotropies in the arrival directions and the exact shape of the suppression. Current technologies can reach one order of magnitude more in the number of observed events at the highest energies, which seems not enough to firmly detect anisotropies or to probe new physics. New technologies are needed and future space observatories, with improved photon detection techniques, like JEM-EUSO, OWL  or Super-EUSO, promise a new era in the physics of UHECR.

\end{document}